\documentclass[a4paper,fleqn]{cas-dc}

\usepackage[numbers,sort&compress]{natbib}

\usepackage{upgreek}
\usepackage{bm}
\def\tsc#1{\csdef{#1}{\textsc{\lowercase{#1}}\xspace}}
\tsc{WGM}
\tsc{QE}
\tsc{EP}
\tsc{PMS}
\tsc{BEC}
\tsc{DE}

\begin{document}
\let\WriteBookmarks\relax
\def\floatpagepagefraction{1}
\def\textpagefraction{.001}

\shorttitle{Universal angled shear wave identity for soft solids}

\shortauthors{H. Berjamin et al.}

\title [mode = title]{Universality of the angled shear wave identity in soft viscous solids}  




%
\author[1]{Harold Berjamin}[orcid=0000-0001-6637-8485]

\cormark[1]


\ead{harold.berjamin@universityofgalway.ie}


\credit{Conceptualization, Methodology, Formal analysis, Writing}

\affiliation[1]{organization={School of Mathematical and Statistical Sciences, University of Galway},
    addressline={University Road}, 
    city={Galway},
    country={Ireland}}

\author[2]{Artur L. Gower}[orcid=0000-0002-3229-5451]

\ead{arturgower@gmail.com}

\credit{Conceptualization, Methodology, Formal analysis, Writing}

\affiliation[2]{organization={Department of Mechanical Engineering, University of Sheffield},
    addressline={University Road}, 
    city={Sheffield},
    country={UK}}

\cortext[cor1]{Corresponding author}

\begin{abstract}
Mechanical stress within biological tissue can indicate an anomaly, or can be vital of its function, such as stresses in arteries. Measuring these stresses in tissue is challenging due to the complex, and often unknown, nature of the material properties. Recently, a method called \emph{the angled shear wave identity} was proposed to predict the stress by measuring the speed of two small amplitude shear waves. The method does not require prior knowledge of the material's constitutive law, making it ideal for complex biological tissues. We extend this method, and the underlying identity, to include viscous dissipation, which can be significant for biological tissues. 
To generalise the identity, we consider soft viscoelastic solids described by a generalised Newtonian viscous stress, and account for transverse isotropy, a feature that is common in muscle tissue, for instance. We then derive the dispersion relationship for small-amplitude shear waves superimposed on a large static deformation. Similarly to the elastic case, the identity is recovered when the stress in the material is coaxial with the transverse anisotropy.
A key result in this paper is that to predict the stress in a viscous material one would need to measure the wave attenuation as well as the wave speed.
The case of viscoelastic materials with memory is also discussed.
\end{abstract}




\maketitle

\section{Introduction}

The function of most materials, and durability, is intimately linked to the level of mechanical stress within the material. For instance, in civil engineering, structures are designed so that internal forces remain below the elastic limit (that is, the yield stress), while also counter-acting the effects of gravity \citep{megson2019structural}. Similarly, mechanical stress plays a major role in biomechanics, such as in the study of bone fracture \citep{fung93}.

Despite its importance, measuring the internal stresses, in a non-destructive way, has proven very challenging. Early attempts to determine residual stress in metals using ultrasonic waves rely on the acousto-elasticity theory, which connects changes in the wave speed to a level of applied stress and some of the material parameters \citep{crecraft67}. This approach can be adapted to modern imaging techniques based on ultrasound, see for instance \citet{bied21}. However, using these techniques to measure the stress has proven difficult, as they require prior knowledge of the material parameters \citep{li20}.

To develop a way to measure the stress, without needing to know the material constants, we can use \emph{universal relations} \citep{truesdell61}. In mechanics, universal relations are mathematical formulas that connect a material's stress to deformation and motion, without involving the material's parameters explicitly. Instead, these relations hold for all materials within a certain class \citep{rivlin00}. For this reason, universal relations are useful for the design of experimental procedures aimed at testing a large range of materials.

In terms of wave propagation, there exists universal relations that connect the wave speeds to the stress in the material, regardless of the material parameters. Examples include Ericksen's identities providing the speed of compression and shear waves in elastic solids \citep{truesdell61}, and an identity by \citet{man87}, which links the stress to the speed of four different shear waves. In both cases, the superimposed waves are assumed non dispersive, i.e., their speed does not depend on frequency. Therefore, these identities do not apply to many soft solids which are inherently viscoelastic, hence dispersive \citep{holzapfel00,volokh2016mechanics}.

\begin{figure}
    \centering
    \includegraphics{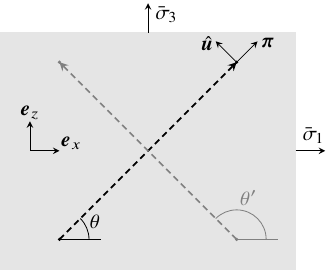}
    \caption{Sketch of the angled shear wave identity in a rectangular cuboid maintained in a state of static stress (sectional view for constant $y$). The selected shear waves propagate along two directions with in-plane polarisation. \label{fig:Config}}
\end{figure}

Named `\emph{angled shear wave identity}', one simple universal relation that connects stress to wave speeds takes the form \citep{li20}
\begin{equation}
    \rho c^2 - \rho c^{\prime 2} = (\bar{\sigma}_1 - \bar{\sigma}_3) \cos (2\theta) ,
    \label{Identity}
\end{equation}
where the mass density $\rho$ is a constant.
This identity connects the difference $\bar{\sigma}_1 - \bar{\sigma}_3$ between the normal stresses along $x$ and $z$ to the speed $c$, $c'$ of only two shear waves, see Figure~\ref{fig:Config}. These waves are both polarised in the same plane $(\bm{e}_x, \bm{e}_z)$ and propagate at specific angles $\theta$ and $\theta' = \pi/2\pm\theta$. Here too, the superimposed waves were assumed non dispersive.

In \citet{zhang23}, the method was applied to the measurement of stress in hydrogel and in skeletal muscle, without the knowledge of the constitutive parameters of the materials. In the supplementary material, the authors study the applicability of the method to transversely isotropic elastic media such as muscle tissue \citep{murphy13}, whose anisotropy derives from the orientation of its fibres. \citet{zhang23} prove that the universal relation still holds as long as the fibres are aligned with one of the principal stresses. Similarly \citet{mukherjee2022representing} demonstrated that the identity \eqref{Identity} holds for compressible elastic solids, with texture anisotropy due to fibres or other structural components, if the texture anisotropy is aligned with the initial stress, and if both the texture anisotropy and stress are small (in comparison to the Lam\'e constants of the material).  

The main aim of this work is to extend this angled shear wave identity to a larger class of materials that are dispersive and incompressible. Examples of materials include soft solids such as hydrogels and biological tissues. The theory of non-linear viscoelasticity offers a highly suitable modelling framework to achieve this, cf. \citet{destrade09}. In this framework, harmonic plane waves no longer propagate with constant waveform and speed. Instead, due to dispersion, waves are attenuated as they propagate, and their speed depends on the frequency \citep{carcione15}. Since the above universal relations were obtained in the elastic case, it remains unclear how they apply to viscoelastic materials.

In the present letter, we consider a general model of soft viscoelastic solid, and the particular case of Kelvin--Voigt solids with Newtonian viscosity (Section~\ref{sec:Constitutive}). These theories account for incompressibility, nonlinear elasticity, viscosity, and transverse isotropy. To analyse the properties of mechanical waves that propagate in such a material subjected to a homogeneous pre-stress, we then proceed with an acousto-elastic linearisation, or small-on-large (Section~\ref{sec:SoL}). The results are used to prove that the angled shear wave identity \eqref{Identity} still holds in viscoelastic solids, as long as the phase velocity is replaced by a complex wave speed which accounts for attenuation (Section~\ref{sec:Identity}). To reach this result, we made the following assumptions:
\begin{enumerate}
    \item the material is incompressible, viscoelastic of differential type, and transversely isotropic;
    \item there exists a dissipation potential from which the viscous stress contribution is derived;
    \item the material is subjected to tri-axial stress, which is aligned with the directions of the transverse anisotropy.
\end{enumerate}
Generalisation to isotropic materials with memory is addressed in the Appendix~\ref{app:Gen}.

\section{Constitutive theory}\label{sec:Constitutive}

Let us introduce the deformation gradient tensor $\bm{F} = \partial \bm{x}/\partial \bm{X}$ which links the current position of a point on a material $\bm x$,  after a deformation, to its initial position $\bm X$ in the reference configuration, before the deformation. The material is assumed \emph{incompressible}, which means that the volume does not change during any deformation. Thus, the determinant of $\bm F$ satisfies $\det \bm F = 1$, and the mass density $\rho$ is constant \citep{holzapfel00}.

According to our assumptions, the Cauchy stress tensor $\bm{T}$ can depend on the elastic deformation, deformation rate, and direction of anisotropy. Here, we decompose the stress as the sum of an elastic contribution $\bm{T}^\text{e}$ and of a viscous one, $\bm{T}^\text{v}$, that is,
\begin{equation}
    \bm{T} = -p\bm{I} + \bm{T}^\text{e} + \bm{T}^\text{v},
    \label{Constitutive}
\end{equation}
where $\bm I$ is the identity tensor. The stress $-p\bm{I}$ is a reaction to enforce the incompressibility constraint. The conservation of angular momentum requires that $\bm{T}$ is symmetric, i.e., equal to its transpose ${\bm T}^\text{T}$.

The expression of the elastic stress is deduced from the strain energy function $W^\text{e}$ as follows \citep{holzapfel00},
\begin{equation}
    \bm{T}^\text{e} = 2\bm{F}\frac{\partial W^\text{e}}{\partial \bm C}\bm{F}^\text{T},
    \label{StressE}
\end{equation}
where $W^\text{e}$ depends on $\bm F$ only through $\bm C$, and
\begin{equation}
    \bm{C} = \bm{F}^\text{T}\bm{F} , \qquad 
    \bm{B} = \bm{F}\bm{F}^\text{T},
    \label{Strain}
\end{equation}
are the right and left Cauchy--Green deformation tensors. Due to incompressibility, the symmetric tensors \eqref{Strain} have unit determinant.

To account for anisotropic materials, we consider a unit vector $\bm n$ which defines the direction of anisotropy in the reference configuration, while $\bm m = \bm F \bm n$ is a vector in the direction of anisotropy in the current configuration. Now by assuming that $W^\text{e}$ depends on only $\bm C$ and $\bm n$, we have that
\begin{equation}
    W^\text{e} := W^\text{e}(\bm{C}, \bm{N}),
    \label{We}
\end{equation}
where the term $\bm{N} = \bm{n}\otimes \bm{n}$ is used instead of $\bm n$ so that swapping $\bm n$ with $-\bm n$ does not change the energy $W^\text{e}$. Likewise, we define $\bm M = \bm{m}\otimes \bm{m}$.

From the assumption \eqref{We}, it can be deduced that $W^\text{e}$ depends on only ten scalar invariants of the tensors $\bm C$, $\bm N$ \citep{spencer1971part,spencer2014continuum}. When further using $\det \bm C = 1$ and the fact that $\bm n$ is a unit vector, these invariants reduce to \citep{murphy13}
\begin{equation}
    \begin{aligned}
     I_1 &= \text{tr}(\bm{C}), \; & 
     I_2 &= \text{tr}(\bm{C}^{-1}) , \\
     I_4 &= \text{tr}(\bm{C}\bm{N}), \; & 
     I_5 &= \text{tr}(\bm{C}^2\bm{N}).
    \end{aligned}
    \label{InvarI}
\end{equation}
The first two scalar functions $I_i$ are the principal invariants of the deformation \eqref{Strain}, whereas the latter ones are related to anisotropy.

To describe the stress due to viscosity ($\bm{T}^\text{v}$), we need to introduce the rate of strain $\dot{\bm C}$ and the symmetric part of the Eulerian velocity gradient $\bm{D} = \frac12 \bm{F}^{-\text{T}}\dot{\bm C}\bm{F}^{-1}$, where the overdot denotes the material time derivative. Due to incompressibility, $\det \bm C = 1$, and as a consequence \citep{destrade09}, we have $\text{tr} \, \bm{D} = 0$.

The viscous stress is deduced from the dissipation potential $W^\text{v}$ such that
\begin{equation}
    \bm{T}^\text{v} = 2\bm{F}\frac{\partial W^\text{v}}{\partial \dot{\bm C}}\bm{F}^\text{T},
    \label{StressV}
\end{equation}
where $W^\text{v}$ can depend on the elastic deformation, deformation rate, and direction of anisotropy $\bm n$, in the form \citep{anssari17}
\begin{equation}
    W^\text{v} = W^\text{v}(\bm{C}, \dot{\bm C}, \bm{N}) .
    \label{Wv}
\end{equation}
The viscous stress \eqref{StressV} vanishes in equilibrium ($\dot{\bm C} = \bm{0})$, where the total stress \eqref{Constitutive} is purely elastic.

From the form~\eqref{Wv}, it can be deduced that $W^\text{v}$ depends only on seventeen scalar invariants of the tensors $\bm C$, $\dot{\bm C}$, $\bm N$, see for instance \citet{merodio06} and \citet{anssari17}, two of which can be eliminated by virtue of the incompressibility property \citep{destrade09}. As pointed out by \citet{coco23}, it seems unpractical to consider such a general theory as it leads to lengthy mathematical expressions.

Although it is not absolutely needed here, we will present a much simpler theory for practical use, namely the transversely isotropic viscous model by \citet{spencer04}. This constitutive model is based on an expression of the dissipation potential in terms of the tensors $\bm D$ and $\bm{A} = \bm{a}\otimes \bm{a}$ only, where $\bm{a} = \bm{m}/\|\bm{m}\|$ is a unit vector defining the current direction of anisotropy, in the deformed configuration (as opposed to the initial direction of anisotropy $\bm{n} = \bm{F}^{-1}\bm{m}$). Thus, the symmetric tensor $\bm{A} = \bm{M}/I_4$ is idempotent with unit trace. With this assumption, the dissipation potential $W^\text{v}$ depends on only ten scalar invariants of $\bm D$, $\bm A$, six of which can be eliminated due to the properties of $\bm A$ and incompressibility.

Next, a Kelvin--Voigt model can be derived, where the viscous stress $\bm{T}^\text{v}$ is chosen to be linear with respect to $\bm D$. With this further simplification, the dissipation potential is now function of $W^\text{v} := W ^\text{v}(J_2, J_4, J_5)$ only, with the invariants defined as
\begin{equation}
    \begin{aligned}
    J_2 &= \text{tr}(\bm{D}^2) = \tfrac14\, \text{tr} \big( (\dot{\bm C}\bm{C}^{-1})^2 \big) , \\
    J_4 &= \text{tr}(\bm{D}\bm{A}) = \tfrac12\, \text{tr}(\dot{\bm C}\bm{N}) / I_4 , \\
    J_5 &= \text{tr}(\bm{D}^2\bm{A}) = \tfrac14\, \text{tr}( \dot{\bm C}\bm{C}^{-1} \dot{\bm C}\bm{N})/I_4 .
    \end{aligned}
    \label{InvarJ}
\end{equation}
Here, we have eliminated those invariants that are cubic in the strain rate, because they lead to non-Newtonian viscosity theories.

We can now deduce a reduced form for the stress tensors. As $W^\text{e}$ depends only on the invariants~\eqref{InvarI}, and $W^\text{v}$ depends only on the invariants~\eqref{InvarJ}, we can calculate the stress tensors \eqref{StressE} and \eqref{StressV} by using the chain rule,
\begin{equation}
    \begin{aligned}
    {\bm T}^\text{e} &= \alpha_1 \bm{B} - \alpha_2 \bm{B}^{-1} + \alpha_4\bm{M} + \alpha_5 (\bm{B}\bm{M} + \bm{M}\bm{B}) ,\\
    {\bm T}^\text{v} &= \beta_2 \bm{D} + \tfrac12 \beta_4 \bm{A} + \tfrac12 \beta_5 (\bm{D}\bm{A} + \bm{A}\bm{D}) ,
    \end{aligned}
    \label{Stress}
\end{equation}
where $\alpha_i = 2\, {\partial W^\text{e}}/{\partial I_i}$ and $\beta_i = 2\, {\partial W^\text{v}}/{\partial J_i}$. The isotropic theory is recovered for $\alpha_4$, $\alpha_5$, $\beta_4$, $\beta_5$ equal to zero.

In general, the elastic moduli $\alpha_i$ (in Pa) are functions of all the selected invariants $I_i$ in Eq.~\eqref{InvarI}. Thus, we observe that the tensor $\bm{M} = I_4 \bm{A}$ can be replaced with $\bm{A}$ in the expression of the elastic stress \eqref{Stress}\textsubscript{1}, provided that the coefficients $\alpha_4$, $\alpha_5$ are given a suitable redefinition. 

Because we assumed that the stress is linear in $\bm D$, the viscosities $\beta_i$ (in Pa{\,}s) are no longer arbitrary functions of the invariants $J_i$ of Eq.~\eqref{InvarJ}. In fact, $\beta_2$ and $\beta_5$ must be constants, whereas $\beta_4$ can be linear in $\bm D$. Thus, we set $\beta_4 = 2 \eta_4 J_4$ with $\eta_4$ constant. With these choices, Eq.~\eqref{Stress}\textsubscript{2} recovers exactly the transversely isotropic Newtonian stress proposed by \citet{spencer04}, Eq.~(34) therein.\footnote{The viscous stress proposed by \citet{coco23} differs from the present one \eqref{Stress}\textsubscript{2} due to substitution of the tensor $\bm A$ by $\bm M$ therein.}

According to the principles of thermodynamics, the dissipation per unit volume of material must remain non-negative to ensure physically admissible behaviour \citep{maugin99}. Here, the dissipation reads
\begin{equation}
    \text{tr}(\bm{T}^\text{v}\bm{D}) = \text{tr}\bigg(\frac{\partial W^\text{v}}{\partial \dot{\bm C}}\dot{\bm C}\bigg) = \beta_2 J_2 + \tfrac12\beta_4 J_4 + \beta_5 J_5 \geq 0 ,
    \label{Dissipation}
\end{equation}
with $\tfrac12\beta_4 J_4 = \eta_4 J_4^2$.  By substituting equation~\eqref{InvarJ} into the above, and using that 
$\bm{T}^\text{v}$ is linear in $\bm{D}$, it can be shown that $\text{tr}(\bm{T}^\text{v}\bm{D})$ is a homogeneous function of degree two with respect to $\dot{\bm C}$, and that the above expression equals $2 W^\text{v}$. Given that the tensor $\bm{D}^2$ is positive semi-definite, we have that the invariants $J_2$, $J_5$ are non-negative. Therefore, requiring that the viscosities $\beta_2$, $\eta_4$, $\beta_5$ to be non-negative is sufficient to ensure thermodynamic consistency.


\section{Acousto-elasticity}\label{sec:SoL}

\subsection{General theory}

In what follows, we linearise the equations of motion with respect to a pre-deformed state, with the goal of studying the propagation of infinitesimal shear waves in a material with initial stress. 


Let us consider a static pre-deformation $\bm{X} \mapsto \bar{\bm x}$ with the deformation gradient $\bar{\bm F} = \partial \bar{\bm x} / \partial{\bm X}$, see notations in \citet{berjamin24b}. In contrast with the updated direction of anisotropy $\bar{\bm m} = \bar{\bm F}\bar{\bm n}$, the initial direction of anisotropy $\bar{\bm n} = {\bm n}$ is unaffected by the pre-deformation.

The initial deformation creates a static initial stress $\bar{\bm T} = -\bar{p}\bm{I} + \bar{\bm T}^\text{e}$, where $\bar{\bm T}^\text{e}$ is deduced from the constitutive law \eqref{Stress}. To describe how elastic waves couple to this stress, we consider a small displacement $\tilde{\bm u} = \bm{x}-\bar{\bm x}$, then follow the `small-on-large' approach \citep{destrade09}. For this purpose, we introduce the increment of the deformation gradient $\tilde{\bm F} = \bm{H}\bar{\bm F}$ such that $\bm{F} = \bar{\bm F}+\tilde{\bm F}$, where $\bm{H} = \nabla \tilde{\bm u}$ is the incremental displacement gradient (here, the gradient stands for $\partial/\partial \bar{\bm x}$).

The equations of motion are then linearised with respect to $\tilde{\bm u}$, in the absence of body forces. Since the deformation is homogeneous, we arrive at the linearised balance of momentum equation \citep{destrade09}
\begin{equation}
    \rho \ddot{\tilde{\bm u}} = \nabla \cdot \tilde{\bm \sigma} ,
    \qquad
    \tilde{\bm \sigma} = -\tilde p\bm{I} + \tilde{\bm \sigma}^\text{e} + \tilde{\bm \sigma}^\text{v} ,
\end{equation}
in the pre-deformed configuration, where $\tilde{\bm \sigma}$ is the incremental Cauchy stress. The incremental displacement $\tilde{\bm u}$ is divergence-free due to the incompressibility constraint. The scalar $\tilde p$ is the incremental pressure and the tensors $\tilde{\bm \sigma}^\bullet$ are the stress increments.

Direct linearisation of the products in Eq.~\eqref{StressE} yields the elastic stress increment $\tilde{\bm \sigma}^\text{e} = [\tilde{\sigma}^\text{e}_{ij}]$, given by
\begin{equation}
    \tilde{\sigma}^\text{e}_{ij} = \bar{T}^\text{e}_{ik}{H}_{jk} + \mathcal{C}_{ijk\ell} H_{k\ell} ,
    \qquad H_{k\ell} = \tilde{u}_{k,\ell} ,
    \label{IncrE}
\end{equation}
with summation over repeated indices. Here,
\begin{equation} \label{eqn:C}
    \mathcal{C}_{ijk\ell} = \delta_{ik} \bar{T}^\text{e}_{j\ell} + 4\bar{F}_{im}\bar{F}_{jn}\frac{ \overline{\partial^2 W^\text{e}} }{\partial C_{mn} \partial C_{pq}} \bar{F}_{kp}\bar{F}_{\ell q}
\end{equation}
defines the \emph{elasticity tensor} $\mathbb{C} = [\mathcal{C}_{ijk\ell}]$, and $\delta_{ik}$ denotes the Kronecker delta. In this formula, the overbar indicates that the tensor derivatives are evaluated in the pre-deformed equilibrium state where $\bm F = \bar{\bm F}$.

As we consider a homogenous pre-deformation we have that $\bar{\bm T}^\text{e}$ and $\mathbb{C}$ do not depend on space, we then compute the divergence of $\tilde{\bm \sigma}^\text{e}$ to get $\tilde{\sigma}^\text{e}_{ij,j} = \mathcal{C}_{ijk\ell} H_{k\ell,j}$, where indices after the comma indicate spatial differentiation. 
We note in passing that the coefficients $H_{jj,k}$ vanish because of incompressibility, and that the last two indices of $H_{k\ell,j}$ can be interchanged because the order of differentiation can be interchanged. The above expressions are equivalent to that found in the supplementary material of \citet{zhang23} up to conventions and notations used therein.

Similarly, we compute the incremental viscous stress by linearisation of \eqref{StressV}, using the absence of strain rates in the pre-deformed
configuration ($\bar{\dot{\bm C}} = \bm{0}$). This way, we arrive at an expression of the viscous stress increment of the form
\begin{equation}
    \begin{aligned}
    &\tilde{\sigma}^\text{v}_{ij} =  \mathcal{D}_{ijk\ell} H_{k\ell} + \mathcal{V}_{ijk\ell} \dot H_{k\ell}  ,\\
    &\mathcal{D}_{ijk\ell} = 4\bar{F}_{im}\bar{F}_{jn}\frac{ \overline{\partial^2 W^\text{v}} }{\partial \dot C_{mn} \partial C_{pq}} \bar{F}_{kp}\bar{F}_{\ell q} ,\\
    &\mathcal{V}_{ijk\ell} = 4\bar{F}_{im}\bar{F}_{jn}\frac{ \overline{\partial^2 W^\text{v}} }{\partial \dot C_{mn} \partial \dot C_{pq}} \bar{F}_{kp}\bar{F}_{\ell q} ,
    \end{aligned}
    \label{IncrV}
\end{equation}
where $\mathbb{D} = [\mathcal{D}_{ijk\ell}]$ is a stiffness tensor and $\mathbb{V} = [\mathcal{V}_{ijk\ell}]$ is the \emph{viscosity tensor}.

For the transversely isotropic Kelvin--Voigt model described earlier, detailed expressions of the stress increments could be derived directly from Eq.~\eqref{Stress} by linearisation, see for instance \citet{destrade09} for the methodology. In particular, we note that the coefficients of $\mathbb{D}$ are equal to zero in this special case.

\subsection{Harmonic plane waves}

We apply this theory to the case of harmonic plane wave solutions of the form
\begin{equation}
    \tilde{\bm u} = \hat{\bm u}\, \text{e}^{\text{i} \omega (t - \bm{\pi} \cdot \bm{x} /c)} ,
    \qquad \text{i} = \sqrt{-1} \, ,
    \label{Harmonic}
\end{equation}
where $\omega > 0$ is the angular frequency and $\bm{\pi}$ is a unit vector along the direction of propagation. Similar notations are introduced for the incremental pressure and the incremental stresses. If the wavenumber $\kappa$ is complex-valued, then the \emph{complex velocity} $c = \omega/\kappa$ is a complex number too.

In general, the complex velocity differs from the phase velocity $v_\varphi = \omega/\mathfrak{Re}(\kappa)$. More precisely, computation of the real and imaginary parts of $1/c$ yields \citep{carcione15}
\begin{equation}
    \mathfrak{Re}(1/c) = 1/v_\varphi , \qquad
    \mathfrak{Im}(1/c) = -\alpha/\omega ,
    \label{Dispersion}
\end{equation}
where $\alpha = -\mathfrak{Im}(\kappa)$ is the attenuation coefficient. If $v_\varphi$ and $\alpha$ are known at a given frequency $\omega$, then the complex velocity $c = (1/c)^{-1}$ can be retrieved based on the above formulas. In the absence of attenuation $(\alpha=0)$, the complex velocity is equal to the phase velocity.

The constraint of incompressiblity $\nabla \cdot \tilde{\bm u} = 0$ leads to the orthogonality property $\bm{\pi}\cdot \hat{\bm u} = 0$, whereas the linearised equation of motion becomes
\begin{equation}
    \rho c^2 \hat{\bm u} = \text{i} \tfrac{c}{\omega} \hat{\bm \sigma} \bm{\pi} = -\text{i} \tfrac{c}{\omega} \hat{p} \bm{\pi} + {\bm Q} \hat{\bm u} ,
    \label{MotHarmo}
\end{equation}
where
\begin{equation}
    Q_{ij} = \mathcal{A}_{ipjq} \pi_p\pi_q, \qquad \mathbb{A} = \mathbb{C} + \mathbb{D} + \text{i}\omega\, \mathbb{V} ,
    \label{Acoustic}
\end{equation}
defines the \emph{acoustic tensor} $\bm Q$ and the \emph{dynamic tensor} $\mathbb{A}$.
On scalar multiplication of \eqref{MotHarmo} by $\bm{\pi}$, we obtain the harmonic amplitude of the incremental pressure, $\hat p$. Then, substitution in \eqref{MotHarmo} yields the eigenvalue problem
\begin{equation}
    \rho c^2 \hat{\bm u} = (\bm{I} - \bm{\pi}\otimes \bm{\pi}) {\bm Q} (\bm{I} - \bm{\pi}\otimes \bm{\pi}) \hat{\bm u} ,
\end{equation}
whose solutions provide the \emph{dynamic modulus} $\rho c^2$.

For non-trivial solutions, the polarisation vector $\hat{\bm u}$ can be assumed unitary, without loss of generality. Then, the dynamic modulus can be retrieved directly from \eqref{MotHarmo}-\eqref{Acoustic} by means of a scalar product with $\hat{\bm u}$:
\begin{equation}
    \rho c^2 = \mathcal{A}_{ipjq} \hat{u}_i\pi_p\hat{u}_j\pi_q .
\end{equation}

Noting that the second term of $\mathbb C$ from \eqref{eqn:C} is formally similar to $\mathbb D$ and $\mathbb V$, see \eqref{IncrV}, we write
\begin{equation}
    \rho c^2 = \pi_p\bar{T}^\text{e}_{pq}\pi_q + \mathcal{M}_{ipjq} \hat{u}_i\pi_p\hat{u}_j\pi_q , 
    \label{DynMod}
\end{equation}
where $\mathbb{M} = [\mathcal{M}_{ipjq}]$ is defined by
\begin{equation}
    \mathcal{A}_{ipjq} - \delta_{ij}\bar{T}^\text{e}_{pq} =: \mathcal{M}_{ipjq} = \mathcal{M}_{pijq} = \mathcal{M}_{ipqj} . 
    \label{M}
\end{equation}
The minor symmetries of $\mathbb M$ follow from the symmetry of $\bm C$. In general, the equality of mixed partials entails the major symmetry $\mathcal{M}_{ipjq} = \mathcal{M}_{jqip}$ for $\mathbb M$ provided that $\mathbb D$ has this symmetry too, a requirement that is not needed here.

\section{The angled shear wave identity}\label{sec:Identity}

We now move on to the derivation of the angled shear wave identity. For this purpose, we assume that the material is subjected to a static triaxial pre-deformation which is aligned with coordinate axes so that 
\begin{equation}
    \bar{\bm F} = \begin{bmatrix}
    \bar\lambda_1 & 0 & 0\\
    0 & \bar\lambda_2 & 0 \\
    0 & 0 & \bar\lambda_3
    \end{bmatrix} , \qquad
    \bar\lambda_1\bar\lambda_2\bar\lambda_3 = 1 ,
\end{equation}
whose determinant equals one, in agreement with the incompressibility property.
This way, the unit vectors $\bm{e}_X, \bm{e}_Y, \bm{e}_Z$ in the reference configuration are equal to their counterpart $\bm{e}_x, \bm{e}_y, \bm{e}_z$ in the pre-deformed configuration. A picture is provided in Figure~\ref{fig:Config}.

Following similar steps shown in \citet{zhang23} for the purely elastic case, we see that the angled shear wave identity only holds when the direction of anisotropy  $\bm{n}$ is aligned with one of the principal directions of the pre-deformation $\bm{e}_X, \bm{e}_Y$, or $\bm{e}_Z$.
This way, the vector $\bar{\bm m} = \bar\lambda_n {\bm e}_n$ is aligned with $\bm{n} = \bm{e}_n$ (no summation over repeated indices), where we have used the notation $\bm{e}_n = \bm{e}_X$ for $n=1$, etc. Furthermore, the tensor $\bar{\bm M} = \bar\lambda_n^2\, {\bm e}_n \otimes {\bm e}_n$ is diagonal. Thus, the pre-stress $\bar{\bm T} = \text{diag}[\bar\sigma_1,\bar\sigma_2,\bar\sigma_3]$ is determined from the principal stresses $\bar\sigma_i = -\bar{p} + \bar{\sigma}_i^\text{e}$, where the coefficients $\bar{\sigma}_i^\text{e}$ are the diagonal entries of $\bar{\bm T}^\text{e}$, see Eq.~\eqref{Stress}\textsubscript{1}, and there is no viscous stress as the pre-deformation is static. Along with these observations, it is worth mentioning that the tensors $\bar{\bm F}$, $\bar{\bm T}$ and $\bar{\bm M}$ are mutually commuting (i.e., their eigenspaces coincide), an alignment condition introduced in the supplementary material of \citet{zhang23}.

We now choose a direction of propagation $\bm{\pi} = \cos\theta\, \bm{e}_x + \sin\theta\, \bm{e}_z$ for the small amplitude shear wave \eqref{Harmonic}. From the orthogonality property $\bm{\pi}\cdot \hat{\bm u} = 0$, we deduce that non-trivial wave solutions are either polarised along $\hat{\bm u} = \bm{e}_y$ (i.e., orthogonally to the propagation plane), or along $\hat{\bm u} = -\sin\theta\, \bm{e}_x + \cos\theta\, \bm{e}_z$, which belongs to the propagation plane. Similarly to earlier works, we now restrict the study to the latter case (\emph{in-plane} polarisation), but a similar methodology could be used for other wave polarisations.

Under the above assumptions, the small-on-large theory \eqref{DynMod} yields
\begin{equation}
    \begin{aligned}
    \rho c^2 &= \bar{\sigma}_1^\text{e} \cos^2\!\theta + \bar{\sigma}_3^\text{e} \sin^2\!\theta + \upalpha \cos^4\!\theta + \updelta \sin^4\!\theta \\
    &\quad + \upbeta\, (\cos^3\!\theta\sin\!\theta - \sin^3\!\theta\cos\!\theta) + \upgamma \cos^2\!\theta \sin^2\!\theta .
    \end{aligned}
    \label{DispersionAngle}
\end{equation}
where
\begin{equation}
    \begin{aligned}
    \upalpha &= \mathcal{M}_{1313} = \updelta ,\\
    \upbeta &= \mathcal{M}_{1333}+\mathcal{M}_{3331}-\mathcal{M}_{3111}-\mathcal{M}_{1113} ,\\
    \upgamma &= \mathcal{M}_{1111} - \mathcal{M}_{1133} - 2\mathcal{M}_{1313} - \mathcal{M}_{3311} + \mathcal{M}_{3333} .
    \end{aligned}
\end{equation}
In the above, the number of coefficients of $\mathbb M$ is reduced to nine due to the symmetries \eqref{M} and the fact that $\hat{\bm u}$, $\bm\pi$ have no component along $\bm{e}_y$.

In a similar fashion to the elastic case \citep{zhang23}, explicit calculation of the coefficients $\mathcal{M}_{ipjq}$ shows that $\upbeta$ vanishes. In fact, the coefficients of $\mathbb M$ involve tensor derivatives of the form $\partial^2 W^\bullet / \partial \bm{U} \partial \bm{V}$ evaluated at equilibrium ($\bm{C} = \bar{\bm C}$, $\dot{\bm C} = \bm{0}$), with $\bm U$, $\bm V$ equal to $\bm{C}$ or $\dot{\bm C}$. In the most general case, the scalar functions $W^\bullet$ defined in \eqref{We}-\eqref{Wv} depend on fifteen invariants of the form $\text{tr}({\bm C}^{l}\dot{\bm C}^{m}\bm{N}^{n})$ with suitable integer exponents \citep{merodio06}. Evaluation of the above partial derivatives by means of the chain rule then leads to $\upbeta = 0$, due to the fact that $\bar{\bm F}$ and $\bm N$ are diagonal.

Finally, by taking $\upbeta = 0$, and rewriting in terms of the total stress components  $\bar\sigma_i = -\bar{p} + \bar{\sigma}_i^\text{e}$, equation \eqref{DispersionAngle} becomes
\begin{multline}
    \rho c^2 = \bar{\sigma}_1 \cos^2\!\theta + \bar{\sigma}_3 \sin^2\!\theta + \bar p + \upalpha 
    \\
    + (\upgamma-2\upalpha) \cos^2\!\theta \sin^2\!\theta ,
    \label{DispersionAngleFinal}
\end{multline}
whose coefficients do not depend on $\theta$. Hence, for any frequency $\omega$, we recover the \emph{angled shear wave identity} \eqref{Identity}
by subtraction of the dynamic moduli $\rho c^2$ and $\rho c^{\prime 2}$ measured along two angles $\theta$ and $\theta' = \pi/2\pm\theta$, respectively (see Figure~\ref{fig:Config}). This exact formula provides direct access to the stress difference $\bar{\sigma}_1 - \bar{\sigma}_3$ based on measurements of the phase velocity and wave attenuation from which $c$ is deduced.

\paragraph*{Illustration.}
Typically, in most ultrasonic experiments, it is simpler to measure the wave speed than it is to measure the attenuation. Here we illustrate what type of error would be expected when only using the wave speed in a viscoelastic solid to predict the stress when using the identity \eqref{Identity}.

For this example, we consider an isotropic neo-Hookean solid with Newtonian viscosity, i.e., we assume that $\alpha_1$ is constant, and that $\alpha_2$, $\alpha_4$, $\alpha_5$, $\beta_4$, $\beta_5$ equal zero. In this case,
\begin{equation}
    \begin{aligned}
        \mathcal{M}_{1111} &= \mathcal{M}_{3333} = 2
        \mathcal{M}_{1313} = \text{i}\omega \beta_2 , \\
        \mathcal{M}_{1133} &= \mathcal{M}_{3311} = 0 .
    \end{aligned}
    \label{CoeffsSimple}
\end{equation}
Thus, the dispersion relationship \eqref{DispersionAngle} with $\upbeta = 0$ can be rewritten as
\begin{equation}
    c^2 = v_0^2 \left( 1 + \text{i} \xi \right) , \qquad
    \xi =  \frac{\omega\beta_2}{2} \frac{\cos^4\!\theta + \sin^4\!\theta}{\rho v_0^2} ,
    \label{IdentityComplex}
\end{equation}
where $\rho v_0^2 = \bar{\sigma}_1^\text{e} \cos^2\!\theta + \bar{\sigma}_3^\text{e} \sin^2\!\theta$ defines the wave speed $v_0$ in the non-dispersive case ($\xi = 0$).

Using Eq.~\eqref{Dispersion}, we can write the squared phase velocity in the form
\begin{equation}
    v_\varphi^2 = v_0^2 \left( 1 + \tfrac34 \xi^2 + \mathscr{O}(\xi^4) \right) ,
    \label{IdentityPhase}
\end{equation}
where we have used a series expansions for small dispersion $\xi$. Therefore, if only the wave speed $v_\varphi$ was measured and used instead of $c$ in \eqref{Identity}, then the predicted stress would be
\begin{equation}
    \frac{\rho v_\varphi^2 - \rho v_\varphi^{\prime 2}}{\cos(2\theta)} \simeq (\bar{\sigma}_1 - \bar{\sigma}_3) \left( 1 - \tfrac34 \xi \xi' \right) ,
    \label{ViscoErr}
\end{equation}
where $v_\varphi'$, $\xi'$ are deduced from \eqref{IdentityComplex}-\eqref{IdentityPhase} with $\theta$ replaced by $\theta'$. In contrast with the exact formula \eqref{Identity}, the above expression includes an error of the order of $(\omega\beta_2)^2$.

This type of error might explain the mismatch found by \citet{zhang23}, according to whom ``the stress is underestimated when the viscoelasticity comes into play''. In the low frequency range of Figure S7B therein, the estimated stress decreases with increasing values of the frequency, which is coherent with Eq.~\eqref{ViscoErr} and $\xi, \xi' \geq 0$. The above formula might also explain the frequency variations in Figure~6 of \citet{delory23}.

\section{Concluding remarks}\label{sec:Conclu}

We have proved that the angled shear wave identity can be generalised, and still holds, for transversely isotropic soft solids with viscosity when replacing the phase velocity with the complex wave velocity. Therefore, if both the wavespeed and attenuation can be measured, then the identity \eqref{Identity}  provides direct access to the initial stress, without extensive prior knowledge of the mechanical behaviour. However, for the identity to hold, the principal directions of the stress need to be aligned with the principal directions of the fibres or texture anisotropy. 

The method we used to prove the identity in this paper involved a pre-deformation to create the initially stressed state. However we note that this proof could have been done for an arbitrary initial stress, without reference to how the stress was created, by following the approach shown in \citet{mukherjee2022representing} and \citet{gower2017new}, see also \citet{ogden23} for details.

Measuring the complex wave velocity is equivalent to measuring both the wave speed and wave attenuation. Such a measurement could be done with an imaging method, as is used for shear wave elastography \citep{zhang23}. If the attenuation can not be measured, then using only the wave speed in place of the complex wave speed can lead to an approximate prediction of the stress. In the specific example considered here, the corresponding error is a quadratic order of the dispersion parameter.

By following similar steps, we believe the identity \eqref{Identity} could be generalised to more complex constitutive models, including compressible ones \citep{li20,mukherjee2022representing}. A first step would be to apply the present method to orthotropic media with two orthogonal preferred directions. In a second step, frequency-dependent responses that go beyond differential models could be investigated, such as models of fractional viscous stress or Maxwell rheologies based on an integral formulation. In the Appendix~\ref{app:Gen}, we show that the result still holds for some isotropic theories of this kind \citep{berjamin24a,berjamin24b}.

\subsection*{Acknowledgements}

HB has received funding from the European Union's Horizon 2020 research and innovation programme under grant agreement TBI-WAVES -- H2020-MSCA-IF-2020 project No 101023950. ALG is grateful for the support from the European Commission Horizon 2020/H2020 - Shift2Rail.

\printcredits

\appendix
\section{Generalisation to materials with memory}\label{app:Gen}

Let us discuss the case of isotropic materials with memory, whose constitutive behaviour agrees with the fractional viscoelastic theory labelled `Model~A' in \citet{berjamin24b}. In a similar fashion to the example of Section~\ref{sec:Identity}, we consider a neo-Hookean elastic stress. Thus, the stress tensor $\bm{T}^\text{e}$ in \eqref{Stress} depends on a single non-zero parameter, $\alpha_1$, which is constant. For causal motions, the viscous stress takes the form of a time-domain convolution product
\begin{equation}
    \bm{T}^\text{v} = \frac{\alpha_1 \tau^\chi}{\Gamma(1-\chi)} \int_0^t \frac{\bm{F}_{t|s}\bm{D}(s)\bm{F}^\text{T}_{t|s}}{(t-s)^\chi}\text{d} s , 
    \label{Conv}
\end{equation}
where $0<\chi<1$, and $\bm{F}_{t|s}\ = \bm{F}(t)\bm{F}^{-1}(s)$ is the relative deformation gradient from the configuration at time $s>0$ to the configuration at time $t>s$. The coefficient $\tau$ is a characteristic time, and $\Gamma$ denotes the gamma function.

By following the steps above and using Eqs. (29)-(32) of Ref.~\cite{berjamin24b}, we recover the dispersion relationship \eqref{DispersionAngleFinal} with the coefficients $\upgamma = 2 \upalpha = \alpha_1 (\text{i}\omega\tau)^\chi$. Thus, Eq.~\eqref{CoeffsSimple} corresponds to the limit $\chi\to 1$ with $\tau = \beta_2/\alpha_1$. Finally, we have shown that the angled shear wave identity \eqref{Identity} still holds for the present theory. A similar derivation could be carried out for generalised Maxwell rheologies by replacing the singular convolution kernel in \eqref{Conv} with a Prony series, see Eqs. (8)-(10) of \citet{berjamin24a}.

\printcredits

\bibliographystyle{elsarticle-num-names}

\bibliography{biblio}


\end{document}